\documentclass[one column]{revtex4}
\usepackage{hyperref}
\usepackage{pdflscape}
 
\usepackage{subfigure}
\usepackage{epsfig}
 
\usepackage{amsmath,amsfonts}
\begin{document}
 
\title{Lorentz-violating  gaugeon formalism for   rank-2 tensor  theory}
 \author{Sudhaker Upadhyay${}^{a,b}$}
 \email{sudhakerupadhyay@gmail.com; sudhaker@associates.iucaa.in}
\author{Mushtaq B. Shah${}^{c}$}
  \email{mbstheory72@gmail.com } 
   \author{Prince A. Ganai${}^{c}$}
  \email{princeganai@nitsri.com}  
   \affiliation{${}^{a}$Department of Physics, K. L. S. College,  Nawada-805110,  India}
   \affiliation{${}^{b}$Visiting Associate, Inter-University Centre for Astronomy and Astrophysics (IUCAA) Pune, Maharashtra-411007}
 \affiliation{${}^{c}$Department of Physics,
National Institute of Technology,
Srinagar, Kashmir-190006, India }

\begin{abstract}
We develop a BRST  symmetric gaugeon formalism for the Abelian rank-2 antisymmetric tensor
 field in the Lorentz breaking framework. The Lorentz breaking is achieved here by 
 considering a proper subgroup of Lorentz group together with translation. 
 In this scenario, the gaugeon fields together with the standard fields  of the Abelian rank-2 antisymmetric tensor theory get mass. In order to develop the gaugeon formulation for this 
 theory in VSR, we first introduce a set of dipole vector  fields   as a quantum gauge freedom to the action.  In order to quantize the dipole
vector   fields, the VSR-modified gauge-fixing and corresponding ghost action are constructed as the classical action is invariant under a VSR-modified gauge transformation.  Further, we present
a Type I gaugeon formalism for the Abelian rank-2 antisymmetric tensor
 field theory in VSR. The gauge structures of Fock space constructed with the help of BRST charges are also discussed. 
 \end{abstract}
\maketitle

\section{Overview and motivations}
It is well-known that Lorentz symmetry is an essential ingredient of 
 a  highly successful model of 
particle physics known as Standard Model (with gauge group $SU(3) \times
SU(2)  \times U(1)$). 
The evidences  of Lorentz symmetry violation  at high energy are still there in the context of string/M-theory \cite{01,02}  and loop quantum gravity
\cite{03}. Remarkably, the  violation of Lorentz symmetry at high energy helps us to renormalize the non-renormalizable interactions, for instance, two fermion-two scalar vertices and four fermion vertices \cite{003}. 
Two interesting theories have been proposed in search of 
Lorentz symmetry violation.  One of them 
considers Planck-scale effects into account by introducing an invariant Planckian parameter into the theory of special relativity and known as Doubly Special Relativity \cite{04,05,06,07,08}. Another is the Very Special Relativity (VSR) which suggests that
the
laws of physics need not be invariant under the full Lorentz group but rather under a proper
subgroup of it \cite{a}. Since energy and momentum should be  conserved for theory, therefore it is mandatory to include spacetime translation together with the $SIM(2)$ group. 
The semi-direct product of $SIM(2)$ group
 with spacetime translation group is known as $ISIM(2)$ group, which  an $8$-dimensional subgroup of the Poincar\'e group. The terms of action
that are invariant only under this subgroup necessarily break discrete symmetries, however the CPT symmetry is preserved
and many empirical successes of special relativity are still functioned.
  This subgroup when supplemented by any of the discrete symmetries $P$, $T$ or $CP$ enlarges 
 to the full Poincar\'e group.

 The idea of breaking the Lorentz symmetry under VSR framework has been considered widely
in the literature. For
 instance, the generalization of VSR to the physical situation ( de Sitter spacetime), in which a cosmological constant is present, has been made \cite{2}. There, it is shown 
that if there is in addition a breaking of de Sitter symmetry, there are corrections near
the endpoint of the beta decay spectrum proportional to the ratio of the square of the
mass, divided by the product of its momentum times its energy. 
The gaugeon formalism and Sakoda's extension of the gauge freedom of the vector field are investigated in the context of VSR and it is found that the gaugeon modes together with gauge modes become massive   \cite{3}. A possibility is discussed that the symmetries underlying the standard model matter and gauge fields are those of Lorentz, while the event space underlying the dark matter and the dark gauge fields supports the algebraic structure underlying VSR  \cite{4}. The  VSR is generalized  to curved spacetimes  and have been found that
gauging the SIM(2) symmetry does not, in general, provide the coupling to the gravitational background  \cite{5}. The generalization 
of VSR to noncommutative spactimes is also studied where 
noncommutative parameter $\theta_{\mu\nu}$ has light-like
character \cite{6, 7}. The algebra followed by VSR gauge transformations is found as closed
algebra and the actions coupling the gauge field to various matter fields  is also constructed within VSR framework  \cite{8}. One possible way to use the spontaneous symmetry-breaking mechanism to give a flavor-dependent VSR
mass to the gauge bosons is also discussed \cite{8,9}.  An interesting VSR based description is proposed that  VSR plays the
same role for
the field theoretic structure of dark
matter  as special relativity plays for standard model \cite{10}. 
The VSR provides  a new mechanism
for neutrino mass which conserves lepton number without introducing additional sterile
states  \cite{11}. The  Super-Yang-Mills theory in SIM(1)  superspace
 and three dimensional Chern-Simons theory \cite{sud00}  are discussed in such schemes \cite{voh,voh1,voh2}.
The superspace description of Lorentz violating $p$-form gauge theories  is also presented \cite{12}.  The VSR generalization to axion electrodynamics is studied and
it is found that the VSR  effects give a health departure from the usual axion field theory \cite{sudd}.  In spite of such investigations, spontaneous breaking of Lorentz symmetry due to ghost condensation  is also explored \cite{mi,mi1,mi2,sud000}. 
In fact, there are  various models in literature  which discuss BRST symmetry \cite{su001,mf,mf1,mos,bodo} and  Lorentz breaking scheme  \cite{pri, pri1}.

On the other hand, the gaugeon formulation was developed originally for the quantum electrodynamics to settle the issues of renormalization of gauge parameter \cite{13}.
The main idea behind the gaugeon formalism was to introduce the  gaugeon fields to the action which describe the quantum gauge freedom.
  Since the gauge freedom at the quantum level doesn't exist as the quantum action is defined only after  fixing the gauge. 
    By introducing the gaugeon fields, it is shown that  there exists a   gauge freedom even after fixing the gauge. The underlying gauge transformation is
    called as the $q$-number gauge transformation. In this mechanism,  the occurrence of shift in gauge parameter during renormalization  was addressed naturally by connecting theories in two different gauges within the same family by a $q$-number gauge transformation. 
      The gaugeon formalism has been applied, and 
     well studied for various gauge fields, such as, Abelian gauge fields \cite{14,15,16,17,18, 19},
     non-Abelian gauge fields \cite{20,21,22,23}, string theories \cite{24,25}, 
     and gravitational fields \cite{26,27}. Recently, The gaugeon formulation of
      the Lorentz invariant  gauge-fixed and quantized dipole vector  field is studied \cite{ao}. The field-dependent BRST transformation \cite{faiz,faiz1,faiz2,sudu,pr,suf,suf1} and $q$-number gauge transformation   are constructed within gaugeon formalism which help in connection
      of different gauge-fixing terms of the action \cite{sudb,sudu1}. 
     The VSR description of the gauge-fixed and quantized dipole vector (gaugeon) field is not studied yet.
    This provides us an opportunity to bridge this gap by  analyzing the VSR effects on the gauge-fixed and quantized vector gaugeon field.  

In this paper, we present  the BRST quantization of Abelian rank-2 tensor theory in VSR 
framework and define physical Hilbert space under Kugo-Ojima condition. This is achieved by adding 
an appropriate  Lorentz breaking non-local terms to the standard action. 
 In this regard, we find that the Kalb-Ramond fields together with
ghost and ghost of ghost fields get the mass  but this can not be an alternative to
Higgs mechanism as all the fields get same tiny mass. Even the fields become massive, the theory admits a VSR-type gauge transformation and needs quantization which is done via Faddeev-Popov  trick.   In order to assign different mass for different fields
there must occur a spontaneous symmetry breaking.  Recently, in  Ref. \cite{ao}, the massless   gaugeon dipole vector model is studied for Abelian rank-2 tensor theory. However, in Lorentz violating  framework, we find that  in order to discuss
the quantum gauge freedom for Abelian rank-2 tensor theory 
 a dipole vector field    becomes massive under VSR framework. This classical  dipole vector 
 theory also admits a VSR-modified gauge invariance. To remove the superfluous degrees of 
 freedom, we fix the gauge and introduce corresponding ghost terms. The 
 (non-local) Faddeev-Popov action for dipole vector field is invariant under a non-local BRST symmetry. 
The non-local generator for this BRST symmetry is calculated. Furthermore,
we construct the gaugeon action for the Abelian rank-2 tensor theory in VSR framework,
where dipole vector fields play the role of quantum (gaugeon) fields. In order to
have a BRST symmetric gaugeon formalism, we introduced a massive ghost fields corresponding to
gaugeon fields. The non-local BRST symmetric gaugeon action admits a non-local $q$-number
gauge transformation. The form-invariance of action requires a shift in gauge parameter automatically. The gaugeon action in VSR admits various sets of BRST transformation
and consequently various BRST charges exist. The Fock spaces constructed with the help of
   these charges are embedded in the physical Hilbert space. 

The plan of the paper is as following. In Sec. \ref{sec2}, we discuss the standard
BRST quantization of Abelian rank-2 tensor theory in the VSR framework. In Sec. \ref{sec3},
we construct a classical theory for dipole vector in VSR scenario and discuss its dynamics.
The BRST quantization of  dipole vector field theory is provided in section \ref{sec4}. 
In Sec.  \ref{sec5}, we develop BRST symmetric gaugeon formalism for 2-form gauge theory in 
VSR which admits $q$-number gauge transformation. We discuss the gauge structure of Fock space for such theory in VSR.  Finally, we summarize this work in section \ref{sec6}.  
\section{BRST quantization of Abelian rank-2 tensor theory in VSR}\label{sec2}
The main feature of VSR is violation of the Lorentz symmetry.
We briefly review  the relevant subgroups involved in VSR. The   generators 
$T_1= K_1 + L_2$ and $T_2= K_2 - L_1$ form a group, isomorphic to the group of translations in the plane and satisfy following algebra:
\begin{eqnarray}
[T_1,T_2]=0,\ \ [L_3, T_1]=T_2, \ \ [L_3, T_2]=-T_2. 
\end{eqnarray} 
The generator $K_3$ together with $T_1$ and $T_2$ form a group known as $HOM(2)$ and satisfy
following algebra:
\begin{eqnarray}
[K_3,T_1]=T_1,\ \ [K_3, T_2]=T_2.  
\end{eqnarray}
The generators $K_3$ and $L_3$   together with $T_1$ and $T_2$ form $SIM(2)$  group, isomorphic to the four parameter similitude group. We study the Abelian antisymmetric rank-2 tensor gauge theory under the Lorentz breaking but $SIM(2)$-invariant setting.
  
Let us begin  with the $D$-dimensional classical action describing 
Abelian antisymmetric rank-2 tensor gauge field $B_{\mu\nu}$  in VSR framework as \cite{sud}
\begin{eqnarray}
 S_0= \frac{1}{12}\int d^D x\ \tilde H^{\mu\nu\lambda}\tilde H_{\mu\nu\lambda},
  \label{class}
\end{eqnarray}  
 where the 3-form wiggle field strength tensor $\tilde{H}_{\lambda\mu\nu}$ is defined as
\begin{eqnarray}
    \tilde{H}_{\mu\nu\lambda}=  \partial_\mu B_{\nu\lambda}
 + \partial_\nu B_{\lambda\mu}  +\partial_\lambda B_{\mu\nu} - 
 \frac{1}{2}\frac{m^{2}}{n\cdot\partial}n_{\mu}B_{\nu\lambda}- \frac{1}{2}\frac{m^{2}}{n\cdot\partial}
 n_{\nu}B_{\lambda\mu} - \frac{1}{2}\frac{m^{2}}{n\cdot\partial}n_{\lambda}B_{\mu\nu}.
\end{eqnarray}
Here null vector $n_\mu=(1,0,0,-1)$ changes only by a constant factor under boosts in the 
$z$- direction. Therefore, the presence of equal number of null vector   in the numerator and  the denominator of quotient leads to the invariance of quotient
  under $HOM(2)$ and $SIM(2)$.  
The  Wiggle field strength 
$H_{\lambda\mu\nu}$  and, hence, action (\ref{class})   
remain invariant 
under the following  VSR modified  gauge transformation:
\begin{eqnarray}
  \delta B_{\mu\nu}&=&\partial_\mu \theta_\nu -  \partial_\nu \theta_\mu
- \frac{1}{2}\frac{m^{2}}{n\cdot\partial}n_{\mu}\theta_{\nu}+ \frac{1}{2}\frac{m^{2}}{n\cdot\partial}
n_{\nu}\theta_{\mu},\label{gau}
\end{eqnarray}                           
   where $\theta_\mu$   is a vector transformation parameter. The choice of gauge 
   parameter, $\theta_\mu=\partial_\mu \zeta- \frac{1}{2}\frac{m^{2}}{n\cdot\partial}n_{\mu}\zeta$, leads to $ \delta B_{\mu\nu}=0$, this implies that the gauge transformation (\ref{gau}) is reducible. For this theory, the further reducibility  identity does not exist. 
 In order to do BRST quantization,  we need   Faddev-Popov ghosts ($\rho_\mu$ and $\bar\rho_{\mu}$), ghost  of  ghost fields ($\phi$, $\bar\phi$, $d$, $\bar d$ and $\eta$) as the theory is reducible.   
Now, we define the Faddeev-Popov action for Abelian rank-2 tensor field in VSR 
 as following:
 \begin{eqnarray} 
S_\mathrm{FP}&= &\int d^D x\left[  \frac{1}{12}\tilde H^{\mu\nu\lambda}\tilde H_{\mu
\nu\lambda} -\partial^{\mu}B^{\nu}B_{\mu\nu}+\frac{1}{2}\frac{m^2}{n\cdot\partial}n^\mu B^\nu
B_{\mu\nu}-\frac{\alpha_1}{2}B^{\mu}B_{\mu}+B^{\mu}\partial_{\mu}\eta
\right. \nonumber\\
& -&  \frac{1}{2}\frac{m^{2}}{n\cdot\partial} (n\cdot B)
\eta + \partial^{\mu}\bar\phi\partial_{\mu}\phi +\left. m^{2}\bar\phi\phi -i\partial^{\mu}\bar\rho^{\nu}\partial_{\mu}\rho_{\nu} -im^2\bar\rho^\mu \rho_\mu  +i\partial^{\mu}\bar\rho^{\nu}\partial_{\nu}\rho_{\mu} \right.\nonumber\\
& +&\left.  \frac{i}{2}\frac{m^2}{n\cdot\partial}(n\cdot\bar\rho)(\partial\cdot\rho)
+\frac{i}{2}\frac{m^2}{n\cdot\partial}(\partial\cdot\bar\rho)(n\cdot\rho)+\frac{1}{4}\frac{m^4}{(n\cdot\partial)^2}(n\cdot\bar\rho)(n\cdot\rho)+i\rho^\mu\partial_\mu d\right.\nonumber\\
&-&\left. \frac{i}{2}
\frac{m^2}{n\cdot\partial}\rho^\mu n_\mu d+i\partial^\mu \bar d\rho_\mu -
\frac{i}{2}\frac{m^2}{n\cdot\partial}n^\mu \bar d \rho_\mu +i\alpha_2 \bar dd\right],
 \label{quant}                      
\end{eqnarray}
where $\alpha_1$ and $\alpha_2$  are gauge parameters and 
$B_\mu$ is  a Nakanishi-Lautrup type auxiliary  multiplier field.
This  action (\ref{quant}) is 
invariant under the following $SIM(2)$-invariant off-shell nilpotent  BRST transformation (${s}_\mathrm{b} ^2=0$):
\begin{eqnarray}
 {s}_\mathrm{b} B_{\mu\nu} &=& \partial_\mu \rho_\nu -  \partial_\nu \rho_\mu-\frac{1}{2}\frac{m^{2}}{n\cdot\partial}n_{\mu}\rho_{\nu} + \frac{1}{2}\frac{m^{2}}{n\cdot\partial}n_{\nu}\rho_{\mu},    \nonumber\\ 
{s}_\mathrm{b} \rho_\mu &=&  -i \partial_\mu \phi +\frac{i}{2}\frac{m^{2}}{n\cdot\partial}n_{\nu} \phi,\qquad   {s}_\mathrm{b} \phi=0,\nonumber\\ 
{s}_\mathrm{b} \bar\rho_{\mu}&=& iB_\mu,\qquad   {s}_\mathrm{b} B_{\mu} =0,\nonumber\\ 
{s}_\mathrm{b} \bar \phi &=& \bar d,\qquad  {s}_\mathrm{b} \bar d =0,\nonumber\\
 {s}_\mathrm{b} \eta &=& d,  \qquad  
{s}_\mathrm{b} d = 0.\label{br}
\end{eqnarray}
This BRST transformation is important to prove renormalizability of Lorentz-violating  
$SIM(2)$-invariant tensor field theory. 
According to Noether's theorem, it is easy to calculate the conserved charge 
  corresponding  to the above BRST transformation, which is given by 
\begin{eqnarray}
 Q_\mathrm{FP} = \int d^{D-1} x \left[B^\lambda {\overleftrightarrow{\tilde\partial_0}} \rho_\lambda
+ \bar d {\overleftrightarrow{\tilde\partial_0}} \phi + (1-\alpha_2) B_0 d\right],
\label{charge}
\end{eqnarray}
where   
${\overleftrightarrow{\tilde\partial_0}} = {\overrightarrow{\tilde\partial_0}}
 -{\overleftarrow{\tilde\partial_0}}$ with $\tilde\partial_0= \partial_0-\frac{1}{2}\frac{m^2}{n\cdot\partial}n_0$. From the above expression, it is evident that this operator (BRST charge), which implements the BRST symmetry in
 the Hilbert space, is nilpotent. In order to have probabilistic interpretation, 
  all the physical
states must be projected  in the positive definite Hilbert space. Now, this charge helps in defining the physical states in total Hilbert
space of the theory  by 
 annihilating the physical states of the total Hilbert space as following:
\begin{eqnarray}
     Q_\mathrm{FP} |\mathrm{phys}\rangle=0. 
\end{eqnarray}
Now, Euler-Lagrange equations of motion for $B_{\mu\nu}, B_\mu$ and $\eta$ are given, respectively, by
\begin{eqnarray}
   && \partial^\lambda H_{\lambda\mu\nu}
  + \partial _\mu B_\nu - \partial _\nu B_\mu -\frac{1}{2}\frac{m^{2}}{n\cdot\partial}n^{\lambda}H_{\lambda\mu\nu}-\frac{1}{2}\frac{m^{2}}{n\cdot\partial}n_{\mu}B_{\nu}+\frac{1}{2}\frac{m^{2}}{n\cdot\partial}n_{\nu}B_{\mu}=0,  \\
   && \partial^\lambda B_{\lambda \mu} +\partial_{\mu}\eta -\frac{1}{2}\frac{m^{2}}{n\cdot\partial}n^{\lambda}B_{\lambda\mu}-\frac{1}{2}\frac{m^{2}}{n\cdot\partial}n_{\mu} { \eta}- \alpha_1 B_\mu =0,    \label{eqn1} \\
&& \partial^\mu B_\mu -\frac{1}{2}\frac{m^{2}}{n\cdot\partial}n^{\mu}B_{\mu}=0. 
\end{eqnarray}
These field equations further lead to
\begin{eqnarray}
 & & (\Box-m^{2}) B_\mu =0,\nonumber\\
 && (\Box-m^{2}) \eta =0.
\end{eqnarray}
These equations suggest that both the fields $ B_\mu$ and $ \eta$ are massive. 
 The equation (\ref{eqn1}) can be recognized as $SIM(2)$-invariant
 the Lorentz-like  gauge condition. 

\section{A  classical dipole vector field theory in VSR}\label{sec3}
In this section, we discuss the classical theory of a dipole vector field  $Y_\mu$,  
which can be considered as a gaugeon field, in the VSR framework. 
 It   explores a  possibility of changing the gauge-fixing parameter within family under
 the so-called 
 $q$-number gauge transformation,   given by 
\begin{eqnarray} 
  B_{\mu\nu} \to {{\tilde{B}}}_{\mu\nu} = B_{\mu\nu} + \tau \left(\partial_\mu Y_\nu - \partial_\nu Y_\mu-\frac{1}{2}\frac{m^{2}}{n\cdot\partial}n_{\mu}Y_{\nu} +\frac{1}{2}\frac{m^{2}}{n\cdot\partial}n_{\nu}Y_{\mu}\right).  \label{qtrans}
\end{eqnarray}
Here $\tau$  is a bosonic transformation parameter.  
The gauge condition in VSR (\ref{eqn1}) changes under 
such a $q$-number gauge transformation as following:
\begin{eqnarray}
 \partial ^\mu B_{\mu \nu} -\frac{1}{2}\frac{m^{2}}{n\cdot\partial}n^{\mu}  B_{\mu 
 \nu} &+&\tau\left[\Box Y_\nu -\partial^\mu\partial_\nu Y_\mu -m^2 Y_\nu +\frac{1}{2}\frac{m^{2}}{n\cdot\partial}n_\nu(\partial\cdot Y)-\frac{1}{2}\frac{m^{2}}{n\cdot\partial}
 \partial_\nu (n\cdot Y)\right.\nonumber\\
 &+&\left. \frac{1}{4} \frac{m^{4}}{(n\cdot\partial)^2}n^\mu n_\nu Y_\mu
  \right] - (\alpha_1 + \tau) B_\nu + \partial_\nu \eta-\frac{1}{2}\frac{m^{2}}{n\cdot\partial}
 n_{\nu}\eta  =0. 
\end{eqnarray}
It should be noted that the gauge-fixing parameter changes to $\alpha_1 + \tau$. 
In order to write the classical action for the dipole vector (gaugeon) field in VSR, we generalize
 the framework of Froissart model  for dipole scalar field \cite{Froissart}. Thus, the classical action for the dipole vector field theory in VSR
  reads   
\begin{eqnarray}  
   S_\mathrm{D}
   &=&\int d^D x\left[  (\Box -m^2)   Y_\ast^\nu   Y_\nu 
   + \partial^\mu Y_\ast^\nu\partial_\nu Y_\mu - \frac{1}{2}\frac{m^{2}}{n\cdot\partial} \partial^\mu Y^\nu_\ast n_\nu  Y_\mu 
  \right.  \nonumber\\
   &-&\left. 
    \frac{1}{2}\frac{m^{2}}{n\cdot\partial} n^\mu Y^\nu_\ast \partial_\nu  Y_\mu 
   +\frac{1}{4} \left(\frac{m^{2}}{n\cdot\partial}\right)^{2} (n\cdot Y_\ast)( n\cdot Y) 
   - \frac{\varepsilon}{2} Y_\ast^\mu Y_{\ast\mu}\right],  \label{dipole}
\end{eqnarray}
where the sign factor $\varepsilon = \pm 1$  and $Y_{\ast \mu}$ is an auxiliary vector 
field.     This action is invariant under the VSR-modified gauge transformation 
\begin{eqnarray}  
Y_\mu \rightarrow Y_\mu+\partial_\mu \theta-\frac{1}{2}\frac{m^2}{n\cdot\partial}n_\mu \theta,
\end{eqnarray}
where $\theta$ represents an arbitrary scalar function. 

The Euler-Lagrange field equations for the dipole vector fields $Y_{\ast \nu}$ and
$Y_{\nu}$, respectively,  can be calculated from the above action (\ref{dipole}) as follows,
\begin{eqnarray}  
   && (\Box -m^{2})Y_{\nu}-  \partial^\mu\partial_\nu Y_\mu +\frac{1}{2}\frac{m^{2}}{n\cdot\partial} \partial^\mu n_\nu Y_\mu + \frac{1}{2}\frac{m^{2}}{n\cdot\partial} n^\mu \partial_\nu Y_\mu -\frac{1}{4} \left(\frac{m^{2}}{n\cdot\partial}\right)^{2}n^\mu n_\nu Y_\mu
                 - \varepsilon Y_{\ast \nu}=0,   \label{Ym}  \\
   && (\Box -m^{2})Y_{\ast\nu}-  \partial^\mu\partial_\nu Y_{\ast\mu} +\frac{1}{2}\frac{m^{2}}{n\cdot\partial} \partial^\mu n_\nu Y_{\ast\mu} + \frac{1}{2}\frac{m^{2}}{n\cdot\partial} n^\mu \partial_\nu Y_{\ast\mu} -\frac{1}{4} \left(\frac{m^{2}}{n\cdot\partial}\right)^{2}n^\mu n_\nu Y_{\ast\mu}=0.                            \label{Y*m}
\end{eqnarray}
These two field equations reflect the following conditions:
\begin{eqnarray}  
\partial^\nu Y_{\ast \nu}-\frac{1}{2}\frac{m^2}{n\cdot\partial}n^\nu Y_{\ast \nu} =0, \ \ \ \ 
  \left(\Box-m^2\right) Y_{\ast \nu} = 0.      \label{box}
 \end{eqnarray}
Exploiting equations  (\ref{Ym}) and (\ref{box}), we get the equation of motion for field $Y_\mu$   as follows,
\begin{eqnarray}
 (\Box -m^{2})^2Y_{\nu} &-&  \Box\partial^\mu\partial_\nu Y_\mu +\frac{1}{2}\frac{m^{2}}{n\cdot\partial}\Box \partial^\mu n_\nu Y_\mu + \frac{1}{2}\frac{m^{2}}{n\cdot\partial} \Box n^\mu \partial_\nu Y_\mu -\frac{1}{4} \left(\frac{m^{2}}{n\cdot\partial}\right)^{2}\Box n^\mu n_\nu Y_\mu\nonumber\\
 & +&m^2  \partial^\mu\partial_\nu Y_\mu -\frac{1}{2}\frac{m^{4}}{n\cdot\partial} \partial^\mu n_\nu Y_\mu - \frac{1}{2}\frac{m^{4}}{n\cdot\partial} n^\mu \partial_\nu Y_\mu +\frac{1}{4} \left(\frac{m^{3}}{n\cdot\partial}\right)^{2}n^\mu n_\nu Y_\mu =0.  \label{dipole1}
  \end{eqnarray}
  The equations of motion (\ref{box}) and (\ref{dipole1}) suggest that the gaugeon fields for the antisymmetric tensor fields would be a massive dipole fields. Thus, the gaugeon fields 
  for Abelian rank-2 tensor field theory get mass $m$ in VSR framework. 
 \section{BRST quantization  of dipole vector field in VSR}\label{sec4}
Since action (\ref{dipole}) is gauge invariant, so possesses some superficial degrees of freedom.
 In order to  quantize it correctly, we need  to impose gauge-fixing
condition which removes the redundancy in gauge degrees of freedom. 
The essential requirements
for gauge-fixing condition are   (i) it must fix the gauge completely, i.e., there must
not be any residual gauge freedom, and (ii) using the transformations it must be possible to bring
any configuration specified by  gaugeon field into one satisfying the gauge condition.
The gauge-fixing can be achieved by adding the appropriate gauge-symmetry breaking terms to the classical action (\ref{dipole})   as follows, 
\begin{eqnarray}  
S_{\mathrm{DGF}}&=& S_{\mathrm{D}}+ \int d^D x\left[  Y_{\ast}^{ \mu} \partial_\mu Y
 +\partial ^\mu Y_{\ast} Y_\mu -\frac{1}{2}\frac{m^2}{n\cdot\partial}Y_\ast^{\mu}n_\mu Y
-\frac{1}{2}\frac{m^2}{n\cdot\partial}Y_\ast(n\cdot Y)+ \alpha_3 Y_\ast Y\right],   \label{lag}
\end{eqnarray}
where $\alpha_3$ is the  gauge-fixing parameter. Here,  $Y_\ast$ and $Y$ represent
 and scalar multiplier fields. 
 
Since the Fock space corresponding to above gauge-fixed action 
 (\ref{lag}) is not positive definite. In order to make it positive definite, 
 we add the  Faddeev-Popov
ghosts $K_\mu$ and $K_{\ast \mu}$ along with scalar FP ghosts  $K$ and $K_\ast$ to the action, 
which compensate  the determinant due to the gauge-fixing term within functional integral, as follows
\begin{eqnarray}  
 S_\mathrm {DFP}
 &= &  \int d^D x\left[  (\Box-m^2)Y_\ast^\nu Y_\nu +\partial^\mu Y_\ast^\nu\partial_\nu Y_\mu  
+\frac{1}{2}\frac{m^2}{n\cdot\partial}(n\cdot Y_\ast) (\partial\cdot Y)
 +\frac{1}{2}\frac{m^2}{n\cdot\partial}(\partial\cdot Y_\ast)(n\cdot Y)\right.\nonumber\\ 
 &+&\left.
 \frac{1}{4}\left(\frac{m^2}{n\cdot\partial}\right)^2(n\cdot Y_\ast)(n\cdot Y) 
 -\frac{\epsilon}{2}Y^\mu_\ast Y_{\ast\mu} + \partial^\mu Y_\ast Y_\mu -\frac{1}{2}\frac{m^2}{n\cdot\partial}Y_\ast (n\cdot Y)+
 Y_\ast^\mu \partial_\mu Y\right.\nonumber\\
 &-&\left. \frac{1}{2}\frac{m^2}{n\cdot \partial}(n\cdot Y_\ast)Y+\alpha_3 Y_\ast Y 
 +i (\Box-m^2)K_\ast^\nu K_\nu +i\partial^\mu K_\ast^\nu\partial_\nu K_\mu  
+\frac{i}{2}\frac{m^2}{n\cdot\partial}(n\cdot K_\ast) (\partial\cdot K)
\right.\nonumber\\ 
 &+&\left.  \frac{i}{2}\frac{m^2}{n\cdot\partial}(\partial\cdot K_\ast)(n\cdot K)+
 \frac{i}{4}\left(\frac{m^2}{n\cdot\partial}\right)^2(n\cdot K_\ast)(n\cdot K) +i\partial^\mu K_\ast K_
 \mu -\frac{i}{2}\frac{m^2}{n\cdot\partial}K_\ast(n\cdot K)\right.\nonumber\\
 & +&\left. iK_\ast^\mu \partial_\mu K
 -\frac{i}{2}\frac{m^2}{n\cdot\partial} (n\cdot K_\ast)K+i\alpha_3 K_\ast K \right]. \label{brs}
\end{eqnarray}
 The quantum action (\ref{brs}) is  invariant under following nilpotent BRST  transformation:
\begin{eqnarray}  
&&    {s}_\mathrm{b} Y_\mu = K_\mu, \ \ {s}_\mathrm{b} K_\mu =0, \ \ {s}_\mathrm{b} K_{\ast \mu} = iY_{\ast \mu},  \ \ {s}_\mathrm{b} Y_{\ast \mu}=0, \nonumber \\
&& {s}_\mathrm{b} Y= K, \ \ {s}_\mathrm{b} K=0, \ \ {s}_\mathrm{b} K_\ast = iY_\ast, 
  \ \ {s}_\mathrm{b} Y_\ast =0,    \label{brst}                   
\end{eqnarray}
Here we note that the gauge-fixed parts of the action is unphysical as it is BRST-exact and so do not contribute to the physical Hilbert space and can be written 
in terms of gauge-fixing fermion $\Psi$ as follows
\begin{eqnarray}  
  S_\mathrm {DFP} = {s}_\mathrm{b} \Psi, 
  \end{eqnarray}
  where the expression of $\psi$ is given by
  \begin{eqnarray}
  \Psi  &=&i\int d^D x  \left[\partial^\mu K_\ast^\nu(\partial_\mu Y_\nu - \partial_\nu Y_\mu )+m^2 K^\nu_\ast Y_\nu +\frac{1}{2}\frac{m^2}{n\cdot\partial}\partial^\mu K_\ast^\nu n_\nu Y_\mu +\frac{1}{2}\frac{m^2}{n\cdot\partial}n^\mu K_\ast^\nu \partial_\nu Y_\mu \right.\nonumber\\
  &-&\left. \frac{1}{4}\left(\frac{m^2}{n\cdot\partial}\right)^2(n\cdot K_\ast)(n\cdot Y)+
     \frac{\epsilon}{2}K^\mu_\ast Y_{\ast\mu} -\partial^\mu K_\ast Y_\mu-K^\mu_\ast\partial_\mu Y +\frac{1}
     {2}\frac{m^2}{n\cdot\partial}K_\ast(n\cdot Y) \right. \nonumber\\
     &+&\left. \frac{1}{2}\frac{m^2}{n\cdot\partial}(n\cdot K_\ast)Y- 
     \alpha_3 K_\ast Y \right].
\end{eqnarray}
Now, using Noether's theorem,  we calculate the conserved charge ($ Q_\mathrm{DFP}$) corresponding to the BRST
transformation (\ref{brst}). This is given by
\begin{eqnarray}
 Q_\mathrm{DFP}=\int  d^{D-1} x \left[Y_{\ast \mu} {\overleftrightarrow{\tilde\partial _0}} K^\mu +(1-\alpha_3)(  Y_{\ast 0} K - Y_\ast K_0)\right]. \label{chr}
\end{eqnarray}
One can check that this charge is nilpotent. This charge helps to define the
physical Hilbert space out of total Hilbert space under Kugo-Ojima condition. 
\section{Gaugeon formalism of 2-form theory in VSR} \label{sec5}
In this section, we discuss the gaugeon formalism for the Abelian antisymmetric 
tensor field $B_{\mu\nu}$ in VSR. It is important to study the gaugeon formalism 
as the renormalized
  gauge parameter  appears naturally  and also this formalism  connect  two different gauges within the same family by the quantum gauge transformation. 
The gaugeon action for the Abelian 2-form gauge theory in VSR framework is given by
\begin{eqnarray}  
  S &= & S_\mathrm{FP}(\alpha_1=0, \alpha_2)+ S_\mathrm{DFP}
  (\alpha_3=\alpha_2)+\int d^D x\left[ \frac{\varepsilon}{2}Y_\ast^\mu Y_{\ast\mu}-\frac{\varepsilon}{2}(Y_{\ast}^\mu + a B^\mu )(Y_{\ast\mu} + aB_\mu)\right],\nonumber\\
&= & \int d^D x\left[ \frac{1}{12}\tilde H^{\mu\nu\lambda}\tilde H_{\mu
\nu\lambda} -\partial^{\mu}B^{\nu}B_{\mu\nu}+\frac{1}{2}\frac{m^2}{n\cdot\partial}n^\mu B^\nu
B_{\mu\nu} +B^{\mu}\partial_{\mu}\eta
-\frac{1}{2}\frac{m^{2}}{n\cdot\partial} (n\cdot B)
\eta +\partial^{\mu}\bar\phi\partial_{\mu}\phi  \right. \nonumber\\
& +&   \left. m^{2}\bar\phi\phi -i\partial^{\mu}\bar\rho^{\nu}\partial_{\mu}\rho_{\nu} -im^2\bar\rho^\mu \rho_\mu  +i\partial^{\mu}\bar\rho^{\nu}\partial_{\nu}\rho_{\mu} +\frac{i}{2}\frac{m^2}{n\cdot\partial}(n\cdot\bar\rho)(\partial\cdot\rho)
+\frac{i}{2}\frac{m^2}{n\cdot\partial}(\partial\cdot\bar\rho)(n\cdot\rho)\right.\nonumber\\
& +&\left. \frac{1}{4}\frac{m^4}{(n\cdot\partial)^2}(n\cdot\bar\rho)(n\cdot\rho)+i\rho^\mu\partial_\mu d-\frac{i}{2}
\frac{m^2}{n\cdot\partial}\rho^\mu n_\mu d+i\partial^\mu \bar d\rho_\mu -
\frac{i}{2}\frac{m^2}{n\cdot\partial}n^\mu \bar d \rho_\mu +i\alpha_2 \bar dd\right.\nonumber
\\
&+&\left.   (\Box-m^2)Y_\ast^\nu Y_\nu +\partial^\mu Y_\ast^\nu\partial_\nu Y_\mu  
+\frac{1}{2}\frac{m^2}{n\cdot\partial}(n\cdot Y_\ast) (\partial\cdot Y)
 +\frac{1}{2}\frac{m^2}{n\cdot\partial}(\partial\cdot Y_\ast)(n\cdot Y)+ \partial^\mu Y_\ast Y_\mu\right.\nonumber\\ 
 &+&\left.
 \frac{1}{4}\left(\frac{m^2}{n\cdot\partial}\right)^2(n\cdot Y_\ast)(n\cdot Y) 
-\frac{\varepsilon}{2}(Y_{\ast}^\mu + a B^\mu )(Y_{\ast\mu} + aB_\mu) -\frac{1}{2}\frac{m^2}{n\cdot\partial}Y_\ast (n\cdot Y)+
 Y_\ast^\mu \partial_\mu Y\right.\nonumber\\
 &-&\left. \frac{1}{2}\frac{m^2}{n\cdot \partial}(n\cdot Y_\ast)Y+\alpha_2 Y_\ast Y 
 +i (\Box-m^2)K_\ast^\nu K_\nu +i\partial^\mu K_\ast^\nu\partial_\nu K_\mu  
+\frac{i}{2}\frac{m^2}{n\cdot\partial}(n\cdot K_\ast) (\partial\cdot K)
\right.\nonumber\\ 
 &+&\left.  \frac{i}{2}\frac{m^2}{n\cdot\partial}(\partial\cdot K_\ast)(n\cdot K)+
 \frac{i}{4}\left(\frac{m^2}{n\cdot\partial}\right)^2(n\cdot K_\ast)(n\cdot K) +i\partial^\mu K_\ast K_
 \mu -\frac{i}{2}\frac{m^2}{n\cdot\partial}K_\ast(n\cdot K)\right.\nonumber\\
 & +&\left. iK_\ast^\mu \partial_\mu K
 -\frac{i}{2}\frac{m^2}{n\cdot\partial} (n\cdot K_\ast)K+i\alpha_2 K_\ast K    \right],  \label{gaugeon}
\end{eqnarray}
where $a$ denotes the group vector valued gauge-fixing parameter. 
 It should be noted that the gauge-fixing parameter $\alpha_1$ mentioned in Faddeev-Popov action for Abelian rank-2 tensor field  (\ref{quant}) 
can be recognized through the parameter $a$ by
\begin{eqnarray}  
\alpha_1 = \varepsilon a^2.    \label{para}
\end{eqnarray}
Remarkably,   the action (\ref{gaugeon}) 
possesses an extra symmetry, so-called $q$-number gauge transformation, which leaves the action 
form-invariant. 
Under $q$-number gauge transformation fields transform as
\begin{eqnarray}  
   & &
  \delta_q B_{\mu\nu}= \hat B_{\mu\nu}-B_{\mu\nu} = \tau \left(\partial_\mu Y_\nu - \partial_\nu Y_\mu 
   -\frac{1}{2}\frac{m^2}{n\cdot\partial}n_\mu Y_\nu 
   +\frac{1}{2}\frac{m^2}{n\cdot\partial}n_\nu Y_\mu\right),
  \nonumber\\
     &&\delta_q \rho_\mu = \hat \rho_\mu - \hat \rho_\mu = \tau K_\mu,\qquad 
   \delta_q Y_{\ast \mu}= \hat Y_{\ast \mu} - Y_{\ast \mu}= - \tau B_\mu, \quad
  \delta_q B_\mu =  \hat B_\mu - B_\mu =0,  
\nonumber \\  
&&  \delta_q    Y_\mu =  \hat Y_\mu - Y_\mu =0,\quad
    \delta_q  K_{\ast \mu}=  \hat K_{\ast \mu} - K_{\ast \mu} =- \tau \bar\rho_{ \mu},
     \quad   \delta_q\bar\rho_{\mu}=
     {\hat {\bar\rho}}_{\mu}- \bar\rho_{\mu}=0,
 \nonumber \\ &&  \delta_q K_\mu = \hat K_\mu - K_\mu =0, \quad \delta_q {\bar d}  =\hat {\bar d}  - \bar d =0, \quad
   \delta_q  \eta= \hat \eta - \eta = \tau Y ,\nonumber\\
   && \delta_q d= \hat d- d = \tau K, 
 \quad
    \delta_q  Y_\ast = \hat Y_\ast - Y_\ast=0,     \quad 
     \delta_q  K_{\ast } =\hat K_{\ast }-K_\ast  =- \tau \bar d, 
 \nonumber \\ &&
      \delta_q Y=  \hat Y-Y =0,      \quad    \delta_q K= \hat K-K=0,
     \quad    \delta_q\phi = \hat \phi- \phi =0, 
     \quad  \delta_q  \bar \phi =\hat {\bar\phi} -\bar \phi =0, 
       \label{qnum}
\end{eqnarray}
where $\tau$ is an infinitesimal transformation parameter having bosonic nature.  
The form-invariance of the gaugeon action (\ref{gaugeon}) under quantum gauge transformation (\ref{qnum}) leads to the following shift in the gauge-fixing parameter:
 the parameter $\hat a$ is defined as 
\begin{eqnarray}  
       \hat a = a + \tau.
\end{eqnarray}
The remarkable feature of the  $q$-number gauge transformation is that it leaves a 
quantum action (having gauge-fixing terms) in VSR invariant.  
The form-invariance signifies that if we replace the original gauge-fixing parameter 
$a$  by shifted parameter $\hat a$, then  the transformed fields
satisfy  the same field equations as the original fields do. 

The equations of motion  for the $B_{\mu\nu}, B_\nu, \eta, \bar\phi$ and $\phi$ fields  corresponding to
action (\ref{gaugeon}) are given by
\begin{eqnarray}  
  & &\partial ^\lambda \tilde{H}_{\lambda \mu \nu} -\frac{1}{2}\frac{m^2}{n\cdot\partial}  n^\lambda \tilde{H}_{\lambda \mu \nu} + \partial_\mu B_\nu - \partial_\nu B_\mu 
 -\frac{1}{2}\frac{m^2}{n\cdot\partial}n_\mu B_\nu +\frac{1}{2}\frac{m^2}{n\cdot 
\partial}n_\nu B_\mu =0,\\
  &&  \partial^\mu B_{\mu\nu} -\frac{1}{2}\frac{m^2}{n\cdot\partial}n^\mu B_{\mu\nu} + \partial_\nu \eta -\frac{1}{2}\frac{m^2}{n\cdot\partial}n_\nu \eta -\varepsilon a (Y_{\ast \nu} + a B_\nu) =0,   \\
  &&   \partial^\mu B_\mu-\frac{1}{2}\frac{m^2}{n\cdot\partial}(n\cdot B) =0,   \\
  &&       (\Box -m^2)\phi     =  0,   \\
  &&   (\Box -m^2) \bar\phi =0.
\end{eqnarray}
It is obvious here that   both the fields $\phi$ and $\bar\phi$ are massive.
The gaugeon fields satisfy the following field equations:
\begin{eqnarray} 
   &&(\Box-m^2)Y_{\nu} - \partial^\mu   \partial_\nu Y_{\mu} 
  +\frac{1}{2}\frac{m^2}{n\cdot\partial}n_\nu(\partial\cdot Y) +\frac{1}{2}\frac{m^2}{n\cdot\partial}\partial_\nu (n \cdot Y)-\frac{1}{4}\left(\frac{m^2}{n\cdot\partial}\right)^2 n_\nu (n\cdot Y) \nonumber\\
  && -\varepsilon(Y_{\ast\nu}+aB_\nu)
  + \partial _\nu Y-\frac{1}{2}\frac{m^2}{n\cdot\partial}n_\nu Y =0, \\
 &&(\Box-m^2)Y_{\ast\nu} - \partial^\mu   \partial_\nu Y_{\ast\mu} 
  +\frac{1}{2}\frac{m^2}{n\cdot\partial}n_\nu(\partial\cdot Y_\ast)  + \frac{1}{2}\frac{m^2}{n\cdot\partial}\partial_\nu (n \cdot Y_{\ast})-\frac{1}{4}\left(\frac{m^2}{n\cdot\partial}\right)^2 n_\nu (n\cdot Y_\ast) \nonumber\\
  &&  
  + \partial _\nu Y_\ast -\frac{1}{2}\frac{m^2}{n\cdot\partial}n_\nu Y_{\ast} =0. 
  \end{eqnarray}
The anti-ghost and ghost fields satisfy the following equations of motions:
 \begin{eqnarray}
  &&(\Box -m^2) \rho_{\nu}   -\partial^\mu \partial_\nu \rho_\mu  
  +\frac{1}{2}\frac{m^2}{n\cdot\partial}n_\nu(\partial\cdot  \rho) 
  +\frac{1}{2}\frac{m^2}{n\cdot\partial}\partial_\nu(n\cdot \rho) -\frac{1}{4}\left(\frac{m^2}{n\cdot\partial}\right)^2n_\nu (n\cdot  \rho) \nonumber \\
  && +\partial_\nu  
   d   -\frac{1}{2}\frac{m^2}{n\cdot\partial}n_\nu d= 0, \\
    &&(\Box -m^2) \bar\rho_{\nu}   -\partial^\mu \partial_\nu \bar\rho_\mu  
  +\frac{1}{2}\frac{m^2}{n\cdot\partial}n_\nu(\partial\cdot  \bar\rho) 
  +\frac{1}{2}\frac{m^2}{n\cdot\partial}\partial_\nu(n\cdot \bar\rho) -\frac{1}{4}\left(\frac{m^2}{n\cdot\partial}\right)^2n_\nu (n\cdot \bar \rho) \nonumber \\
  && +\partial_\nu    \bar d   -\frac{1}{2}\frac{m^2}{n\cdot\partial}n_\nu\bar d= 0, 
  \end{eqnarray}
 The equations of motion for   anti-ghost and ghost fields corresponding to gaugeon fields,
 respectively, are
  \begin{eqnarray}
      &&(\Box -m^2) K_{\nu}   -\partial^\mu \partial_\nu K_\mu  
  +\frac{1}{2}\frac{m^2}{n\cdot\partial}n_\nu(\partial\cdot K) 
  +\frac{1}{2}\frac{m^2}{n\cdot\partial}\partial_\nu(n\cdot K) -\frac{1}{4}\left(\frac{m^2}{n\cdot\partial}\right)^2n_\nu (n\cdot  K) \nonumber \\
  && +\partial_\nu  
   K  -\frac{1}{2}\frac{m^2}{n\cdot\partial}n_\nu K= 0, \\
    &&(\Box -m^2)\bar K_{\nu}   -\partial^\mu \partial_\nu \bar K_\mu  
  +\frac{1}{2}\frac{m^2}{n\cdot\partial}n_\nu(\partial\cdot  \bar K) 
  +\frac{1}{2}\frac{m^2}{n\cdot\partial}\partial_\nu(n\cdot \bar K) -\frac{1}{4}\left(\frac{m^2}{n\cdot\partial}\right)^2n_\nu (n\cdot  \bar K) \nonumber \\
  && +\partial_\nu  
 \bar K  -\frac{1}{2}\frac{m^2}{n\cdot\partial}n_\nu \bar K= 0.
\end{eqnarray}
The effective gaugeon action for 2-form theory in VSR (\ref{gaugeon}) is invariant   under 
the following nilpotent BRST transformations: 
\begin{eqnarray}
  &&{s}_\mathrm{b} B_{\mu\nu}= \partial_\mu \rho_\nu -  \partial_\nu \rho_\mu-\frac{1}{2}\frac{m^2}{n\cdot\partial}n_\mu \rho_\nu+\frac{1}{2}\frac{m^2}{n\cdot\partial}n_\nu \rho_\mu
 \nonumber\\ 
    &&{s}_\mathrm{b} \rho_\mu= -i \partial_\mu \phi +\frac{i}{2}\frac{m^2}{n\cdot\partial}n_\nu \rho_\mu \phi, \ \ {s}_\mathrm{b} \phi=0,\ \ {s}_\mathrm{b} B_\mu=0, \nonumber\\
  && {s}_\mathrm{b} \bar\rho_{\mu}= iB_\mu, \ \  {s}_\mathrm{b} \bar\phi =\bar d,\ \ 
  {s}_\mathrm{b} \eta = d,\ \ {s}_\mathrm{b} d=0,\nonumber\\
           &&{s}_\mathrm{b} Y_\mu = K_\mu, \ \ {s}_\mathrm{b} K_\mu =0, \ \ {s}_\mathrm{b} K_{\ast 
    \mu} = iY_{\ast \mu},   \ \ {s}_\mathrm{b} Y_{\ast \mu}=0,   \nonumber \\
    &&{s}_\mathrm{b} Y= K, \ \ {s}_\mathrm{b} K=0, \qquad {s}_\mathrm{b} K_\ast = iY_\ast, 
     \ \ {s}_\mathrm{b} Y_\ast =0.         \label{brst2}                   
\end{eqnarray}
Here we see that   the   fields of gaugeon sector form the BRST quartet.
The unphysical parts of the action (\ref{gaugeon}) is BRST exact and can be written in terms of gauge-fixing fermion ($\Psi_g$) as 
\begin{eqnarray}  
 S &=  &\int d^D x\left[  \frac{1}{12}  \tilde H^{\mu\nu\lambda}\tilde H_{\mu\nu\lambda}  \right]+s_\mathrm{b}\Psi_g, \label{gac}
\end{eqnarray}
where 
\begin{eqnarray}  
\Psi_g &=  & i \int d^D x  \left [\partial^\mu \bar\rho^\nu(\partial_\mu B_\nu - \partial_\nu B_\mu) +m^2 \bar\rho^\nu B_\nu  -\alpha_2 K_\ast Y +\frac{1}{2}\frac{m^2}{n\cdot\partial} \partial^\mu \bar\rho^\nu n_\nu B_\mu-\alpha_2 \bar d \eta\right.
 \nonumber\\
&+&\left. 
 \frac{1}{2}\frac{m^2}{n\cdot\partial} n^\mu \bar\rho^\nu \partial_\nu B_\mu +\frac{1}{4}\left(\frac{m^2}{n\cdot\partial}\right)^2 (n\cdot\bar\rho )(n \cdot B)-\bar\rho^\mu\partial_\mu\eta  +\frac{1}{2}\frac{m^2}{n\cdot\partial}(n\cdot \bar\rho )\eta + \partial^\mu \bar\phi  \rho _\mu 
\right. \nonumber \\
&-&\left. \frac{1}{2}\frac{m^2}{n\cdot\partial}\bar\phi(n\cdot \rho) + \frac{\varepsilon}{2} (K_\ast^\mu + a \bar\rho^\mu) ( Y_{\ast \mu } + a B_\mu) + \partial^\mu K_\ast^\nu (\partial_\mu Y_\nu - \partial_\nu Y_\mu) -m^2 K^\mu_\ast Y_\mu\right. \nonumber\\
 &+&\left. \frac{1}{2}\frac{m^2}{n\cdot\partial} n^\mu K_\ast^\nu  \partial_\nu Y_\mu +\frac{1}{2}\frac{m^2}{n\cdot\partial}\partial^\mu K_\ast^\nu  n_\nu Y_\mu -\frac{1}{4}\left(\frac{m^2}{n\cdot\partial}\right)^2(n\cdot Y)(K_\ast\cdot n)-K^\mu_\ast\partial_\mu Y \right. 
 \nonumber\\
&+&\left. \partial^\mu K_\ast Y_\mu+\frac{1}{2}\frac{m^2}{n\cdot\partial}(K_\ast\cdot n)Y -\frac{1}{2}\frac{m^2}{n\cdot\partial}K_\ast(n\cdot Y) \right]. 
\end{eqnarray}
The   invariance of the action (\ref{gac}) under BRST transformation is obvious due to nilpotency of the BRST transformation. 
We calculate the conserved BRST charge $Q_{\rm B}$ with the help of Noethers's theorem
as follows,
\begin{eqnarray}
  Q_\mathrm {B} &=& \int  d^{D-1} x \left[ B^\lambda {\overleftrightarrow{\tilde\partial_0}} \rho_\lambda + \bar d  {\overleftrightarrow{\tilde\partial_0}} \phi + (1-\alpha_2) B_0 d + Y_{\ast}^\lambda {\overleftrightarrow{\tilde\partial _0}} K_\lambda\right.\nonumber\\
  & +&\left.(1-\alpha_2)(  Y_{\ast 0} K - Y_\ast K_0)\right]. \label{chrg}
\end{eqnarray}
The above BRST charge helps to  define the 
Fock space of the system (which is Kernel of BRST charge) in following manner:
\begin{eqnarray}  
\mathcal V _{\rm phys} =  \{ |\Phi \rangle; \, Q_{\rm B} |\Phi \rangle =0\}. 
\label{vec}
\end{eqnarray}
 The BRST transformations 
  (\ref{brst2}) commute  with the  $q$-number gauge transformation (\ref{qnum}).
 This suggests that the the BRST charge is invariant
  under the $q$-number gauge transformations, i.e., 
\begin{eqnarray}  
 \delta_q  Q_\mathrm{B} =  0. 
\end{eqnarray}
Consequently, the fock space (the  physical subspace  of states) $\mathcal V_\mathrm{phys}$  is invariant
under the $q$-number gauge transformation, i.e., 
\begin{eqnarray}  
   \delta_q {\mathcal{V}}_\mathrm{phys}= 0. 
\end{eqnarray}
As a result, the physical Hilbert space of 2-form gauge theory in VSR
$\mathcal H _{\rm phys}  = \mathcal V _{\rm phys}/\mbox{Im} Q$ is also invariant under both the BRST and the  $q$-number  gauge transformations.

There exist many nilpotent symmetry and corresponding charges for the action (\ref{gaugeon}) 
in addition to charge (\ref{chrg}). For instance, these charges are 
\begin{eqnarray}  
  Q_\mathrm{FP} &= &\int  d^{D-1} x  \left[ B^\mu {\overleftrightarrow{\tilde\partial_0}} \rho_\mu + \bar d  {\overleftrightarrow{\tilde\partial_0}} \phi + (1-\alpha_2) B_0 d \right],               \label{charge_b}  \\
 Q_\mathrm{DFP}& =&\int   d^{D-1} x\left[Y_{\ast}^{ \mu} {\overleftrightarrow{\tilde\partial _0}} K_\mu+(1-\alpha_3)( Y_{\ast 0} K - Y_\ast K_0)\right],  \label{vsrch} \\
   Q'_\mathrm{B}& =&\int d^{D-1} x \left[  B^\mu {\overleftrightarrow{\tilde\partial _0}} K_\mu+(1-\alpha_3) B_0 K  \right].    \label{vsrch1}
\end{eqnarray}
The charge $Q_\mathrm{FP}$  is the generator of
 the BRST transformation 
(\ref{br}) while charge
$Q_\mathrm{DFP}$  is the
generator of  the BRST transformation (\ref{brst2}).
The charge $Q'_\mathrm{B}$ (\ref{vsrch1}) generates following BRST  transformation 
${s}'_\mathrm{b}$: 
\begin{eqnarray}  
  &&{s}'_\mathrm{b} B_{\mu\nu} = \partial_\mu K_\nu - \partial_\nu K_\nu -\frac{1}{2}\frac{m^2}{n\cdot\partial}n_\mu K_\nu   +\frac{1}{2}\frac{m^2}{n\cdot\partial}n_\nu K_\mu, \nonumber\\
 && {s}'_\mathrm{b} K_{\ast \mu} = i B_\mu, \qquad {s}'_\mathrm{b} \eta = K,  \qquad {s}'_\mathrm{b} (\text{other fields}) =0.
\end{eqnarray}
These charges satisfy following anticommutation relation among themselves:    
\begin{eqnarray} 
    \{ Q_\mathrm{FP}, Q_\mathrm{DFP} \}
   = \{ Q_\mathrm{DFP},   Q'_\mathrm{(B)} \}
   = \{  Q'_\mathrm{B}, Q_\mathrm{FP} \}
     =0.
\end{eqnarray}
The BRST charge  $Q_\mathrm{FP}$ acts separately for fields of standard formalism sector and 
the BRST charge  $Q_\mathrm{DFP}$ act separately for  fields of gaugeon sector.  
 The net BRST charge  
(\ref{chrg}), which acts on the fields of both the standard formalism sector  and  gaugeon sector, is given by 
\begin{eqnarray}
Q_\mathrm{B}=Q_\mathrm{FP}+Q_\mathrm{DFP}.
\end{eqnarray} 
This charge is the generator of BRST transformation (\ref{brst2}). 

In VSR framework also, we can  define a subspace of states in total Hilbert space $\mathcal V ^{(a)}_\mathrm{phys}$ as  
 \begin{eqnarray}  
   \mathcal V ^{(a)}_\mathrm{phys}=\ker\ Q_\mathrm{FP} \cap \ker\ Q_\mathrm{DFP}
 = \{ |\Phi \rangle ; \, Q_\mathrm{FP}|\Phi \rangle  = Q_\mathrm{DFP} |\Phi \rangle =0 \}.
\end{eqnarray}
Here, the index $(a)$ in  
$\mathcal V^{(a)}_\mathrm{phys}$ 
signifies the  dependence of its definition on the gauge-fixing parameter $a$.
 For such space, the Kugo-Ojima condition, 
  $Q_\mathrm{FP} |\Phi \rangle =0$, removes the  superfluous  
modes of the standard formalism sector. However, the Kugo-Ojima condition, $Q_\mathrm{DFP} |\Phi \rangle =0$, removes the  superfluous  
 modes of the gaugeon sector.  
 It is very obvious now that the space $   \mathcal V^{(a)}_\mathrm{phys} \subset \mathcal V _\mathrm{phys}$. 
 
 Also, the BRST charges $Q _\mathrm{FP}$ and $Q_\mathrm{DFP}$ 
transform under the $q$-number gauge transformation (\ref{qnum}) as follows,
\begin{eqnarray}  
&&\delta_q Q_\mathrm{FP} =   \tau  Q'_\mathrm{B},   \nonumber \\
&&\delta_q Q_\mathrm{DFP} =  - \tau   Q'_\mathrm{B},    \label{vsrch2}
\end{eqnarray}
which insures that the sum of the charges ($Q_\mathrm{B}$) is invariant under $q$-number gauge transformation . 

We also define the subspace  $\mathcal V^{(a)}$ of the total Fock space as
\begin{eqnarray}  
\mathcal V^{(a)} = \ker\ Q_\mathrm{DFP}  =  \{ |\Phi \rangle; \, _\mathrm{DFP}|\Phi \rangle =0\}.  
\end{eqnarray}
This  space coincides with the space of physical dipole vector field for  $\alpha_1=\varepsilon a^2$ gauge. 
 This implies that the  action (\ref{gaugeon}) in $\alpha_1=\varepsilon a^2$ gauge can be written as
\begin{eqnarray}  
  S = S_\mathrm{FP}(\alpha_1=\varepsilon a^2) +i\int dt \{ Q_\mathrm{B}, \varTheta \},
 \label{gaug}
\end{eqnarray}
with  
\begin{eqnarray}
\varTheta &= & \frac{\varepsilon}{2} K_\ast^\mu(Y_{\ast\mu}+ 2a B_\mu)+\partial^\mu K_\ast^\nu 
(\partial_\mu Y_\nu - \partial_\nu Y_\mu)+m^2K_\ast^\nu Y_\nu +\frac{1}{2} \frac{m^2}{n\cdot\partial} \partial^\mu  K_\ast^\nu n_\nu Y_\mu  \nonumber\\
&+& \frac{1}{2} \frac{m^2}{n\cdot\partial} n^\mu  K_\ast^\nu \partial_\nu Y_\mu -
\frac{1}{4} \left(\frac{m^2}{n\cdot\partial}\right)^2(n\cdot K_\ast)(n\cdot Y)-K^\mu_\ast\partial_{\mu}Y
 \nonumber\\
&+& \partial^\mu K_\ast Y_\mu+\frac{1}{2}\frac{m^2}{n\cdot\partial}(n\cdot K_\ast)Y - \frac{1}{2}\frac{m^2}{n\cdot\partial}K_\ast(n\cdot Y)-\alpha_2 K_\ast Y.
\end{eqnarray}
 The first term of action in VSR (\ref{gaug})  refers to the 
action of the standard formalism (\ref{quant}), while the second term 
describes the null operation in the subspace $\mathcal V^{(a)}$ 
 and hence can be ignored in  $\mathcal V^{(a)}$. 

 The same arguments hold for the quantum gauge transformed  
 BRST charges also as for the original BRST charges.
 If  we define the  
subspaces 
$\mathcal V^{(a+\tau)}$ and  
$\mathcal V^{(a+\tau)}_\mathrm{phys}$ annihilated by the $q$-number gauge transformed BRST charges as
\begin{eqnarray}  
  &&\mathcal V^{(a+\tau)} = \ker  (Q_\mathrm{DFP}+\delta_q Q_\mathrm{DFP}), \nonumber \\
  &&\mathcal V^{(a+\tau)}_\mathrm{phys} = \ker   (Q_\mathrm{FP}+\delta_q Q_\mathrm{FP})
 \cap \ker  
  (Q_\mathrm{DFP}+\delta_q Q_\mathrm{DFP}). 
\end{eqnarray}
where $\alpha_1 =\varepsilon (a+\tau)^2$, and   
the corresponding physical subspace is 
$\mathcal V^{(a+\tau)}_\mathrm{phys}$.
Consequently, in VSR scenario also,  the   single Fock space corresponding to BRST invariant 
gaugeon action 
  of 2-form gauge theory    
embeds the Fock spaces of the 2-form gauge theory in
family of linear gauges. 

We would like to comment here that the Type II gaugeon formalism for the present theory can also be developed in VSR framework.
For Type II theory, 
the standard gauge-fixing parameter $\alpha_1$ is expressed as   $\alpha_1=a$ and 
the sign of $\alpha_1$ can also be changed in Type II theory, however, for 
Type I theory
$\alpha_1=\varepsilon a^2$. 
In both cases, the gauge-fixing parameter $a$ 
can be shifted as $\hat a= a+ \tau$ by the $q$-number gauge transformation.

For Type II,  the action in VSR is given by  
\begin{eqnarray}  
 S_\mathrm{II} & =&  S_\mathrm{FP}(\alpha_1=a;\alpha_2) +  S_\mathrm {DFP}(\alpha_3=\alpha_2) 
 +\int d^Dx\left[\frac{\varepsilon}{2}Y_\ast^\mu Y_{\ast \mu} -\frac{1}{2} Y_\ast^\mu B_\mu\right]. \label{diff}
\end{eqnarray}
 Here, the standard gauge-fixing parameter $\alpha_1$ 
can be identified with 
$\alpha_1 = a
$.  
The action (\ref{diff}) remains invariant under the BRST transformations generated by
  the BRST charges (\ref{chrg}), (\ref{charge_b}), (\ref{vsrch}), and (\ref{vsrch1}) too. 
Therefore, the   single Fock space corresponding to Type II  
gaugeon action 
  of 2-form gauge theory   
embeds the Fock spaces of the 2-form gauge theory in
family of linear gauges. 

\section{Concluding remarks}\label{sec6}
  In this paper, we have analysed  the BRST quantization of $SIM(2)$ invariant Abelian rank-2 tensor theory  
framework and define physical Hilbert space under Kugo-Ojima condition. 
The rotational symmetry of full Lorentz group is broken by fixing the null direction. This is achieved by adding 
the appropriate  Lorentz breaking non-local terms to the standard action. 
 In this regard, we find that the Abelian rank-2 tensor fields together with
ghost and ghost of ghost fields satisfy the Proca-type equations and 
get the mass  but this can not be an alternative to
Higgs mechanism as all the fields get same value of  mass. The spontaneous symmetry breaking should take place in order to assign different mass for different fields. We have found that the VSR modified 
action is not invariant under usual gauge transformation rather this is invariant under  a VSR-modified gauge transformation. As the theory possesses the gauge invariance, 
it contains superfluous degrees of freedom. In order to remove these extra degrees of freedom,  we have fixed the gauge which introduced the Faddeev-popov ghost terms. 
The nilpotent BRST transformation and their generator are constructed for the resulting effective action. 
 
   Recently, the massless   gaugeon dipole vector model is studied for Abelian rank-2 tensor theory satisfying full Lorentz group \cite{ao}. Our motivation here is to 
   discuss the $SIM(2)$-invariant generalization of the  gaugeon dipole vector theory. 
  We have observed that  the dipole vector field  gets mass  under the VSR framework, 
  still this classical  dipole vector 
action also admits a VSR-modified gauge invariance. In order to remove the redundant degrees of   freedom due to gauge symmetry, we have chosen a VSR-modified  gauge. This is implemented in the action by adding the suitable non-local gauge-fixing and ghost terms.   
The BRST transformation and their generator are constructed  for the
 (non-local) effective action for dipole vector field.   Furthermore, in order to discuss
the quantum gauge freedom for Abelian rank-2 tensor theory  in VSR,
we have constructed the gaugeon action for the Abelian rank-2 tensor theory,
where dipole vector fields play the role of quantum (gaugeon) fields.
 In order to replace the Gupta-Bleuler type subsidiary condition which removes the unphysical
 gaugeon mode to Kugo-Ojima type subsidiary condition, we have developed 
 a BRST symmetric gaugeon formalism by introducing the massive ghost fields corresponding to
gaugeon fields to the action. We have shown that the $SIM(2)$-invariant BRST symmetric gaugeon action admits a (non-local) $q$-number
gauge transformation. The form-invariance of action requires a shift in gauge parameter which can be identified as the renormalized gauge parameter. We noted here that the BRST transformation  commutes with the $q$-number gauge transformation in VSR also, therefore, 
the physical Hilbert space of 2-form gauge theory   
$\mathcal H _{\rm phys}  = \mathcal V _{\rm phys}/\mbox{Im} Q$ is also invariant under both the BRST and the  $q$-number  gauge transformations.
We have found that there exist various sets of BRST transformation for the gaugeon action in VSR   
and therefore various BRST charges exist. We have shown that the Fock spaces constructed with the help of 
   these charges are embedded into a single physical Hilbert space.


\begin{thebibliography}{99}
\bibitem{01}  V. A. Kostelecky and S. Samuel, Phys. Rev. D 39, 683 (1989).
\bibitem{02}  V. A. Kostelecky and R. Potting, Phys. Rev. D 51, 3923 (1995); Phys. Lett. B 381, 389 (1996).
\bibitem{03}  R. Gambini and J. Pullin, Phys. Rev. D 59, 124021 (1999); J. Alfaro, H. A. Morales-Tecotl and L. F. Urrutia, Phys. Rev. Lett. 84, 2318 (2000).
\bibitem{003}D. Anselmi and M. Halat, Phys. Rev. D 76, 125011 (2007).
\bibitem{04} G. Amelino-Camelia,   Phys. Lett. B 510,  255 (2001).
\bibitem{05}  G.  Amelino-Camelia, Int. J. Mod. Phys. D 11, 35 (2002).
\bibitem{06}  G.  Amelino-Camelia, Nature   418, 34 (2002).
\bibitem{07}J.  Magueijo and L. Smolin, Phys. Rev. Lett.  88, 190403 (2002).
\bibitem{08} J.  Magueijo and L. Smolin, Phys. Rev. D 67, 044017  (2003).

  \bibitem{a}  A. G. Cohen and S. L. Glasow,   Phys. Rev. Lett. 97, 021601 (2006).

 
           

  
 \bibitem{2}   E. Alvarez and R. Vidal,     Phys. Rev. D 77, 127702 (2008).
          
 \bibitem{3}S. Upadhyay and P.K. Panigrahi,   Nucl. Phys. B   915, 168 (2017).     
 \bibitem{4} D. V. Ahluwalia and S. P. Horvath,  JHEP 1011, 078 (2010). 
 \bibitem{5}  W. Mueck,  Phys. Lett.B 670, 95 (2008).
     
 \bibitem{6}  M. M. Sheikh-Jabbari and A. Tureanu,  Phys. Rev. Lett. 101, 261601 (2008).  
 \bibitem{7} S. Das, S. Ghosh and S. Mignemi,   Phys. Lett.  A 375, 3237 (2011).  
 \bibitem{8}  J. Alfaro and V.O. Rivelles,  Phys. Lett. B 734,  239 (2014).    
 \bibitem{9} S. Cheon, C. Lee and S. Lee,  Phys. Lett.  B 679, 73 (2009).  
 \bibitem{10}  S. Masood, M. B. Shah and P. A. Ganai, Int. J. Geom. Methods  Mod. Phys. 15, 1850021 (2018).
 \bibitem{11}  A. G. Cohen and S. L. Glashow, arXiv:hep-ph/0605036.
 \bibitem{sud00}S. Upadhyay, Phys. Rev. D 92, 065027 (2015);  Int. J. Mod. Phys. A 31, 1650112 (2016).
 \bibitem{voh}J. Vohanka and M. Faizal,  Phys. Rev. D 91,   045015 (2015).
 \bibitem{voh1}J. Vohanka and M. Faizal,  Nucl. Phys. B 904, 327 (2016).
 \bibitem{voh2}J. Vohanka and M. Faizal, Eur. Phys. J. C 75, 592 (2015) 
   \bibitem{12} S. Upadhyay, M. B. Shah and P. A. Ganai,   Eur. Phys. J. C 77, 157 (2017).
   \bibitem{sudd}R. Bufalo  and S. Upadhyay, Phys. Lett. B 772, 420 (2017).
   \bibitem{mi}M. Faizal,  J. Phys. A 44 (2011) 402001. 
   \bibitem{mi1}M. Faizal and S. Upadhyay, Phys. Lett. B736, 288 (2014).
   \bibitem{mi2}S. Upadhyay, Int. J.   Mod. Phys. A
  30,   (2015) 1550150.
  \bibitem{sud000}S. Upadhyay, EPL  117 (2017) 11001.
  \bibitem{su001} S. Upadhyay, Prog. Theor. Exp. Phys. 093B06, 1 (2015);
  M. Faizal,  Found. Phys. 41, 270 (2011); M. Faizal and M. Khan, Eur. Phys. J. C 71, 
  1603 (2011).
  \bibitem{mf}M. Faizal,  Phys. Lett. B 705, 120 (2011); Phys. Rev. D 84, 106011 (2011);
  	Mod. Phys. Lett. A 27, 1250075 (2012);  Mod. Phys. Lett. A 27, 1250147 (2012);
  	Comm. Theor. Phys. 58, 704 (2012); JHEP  1301,  156 (2013); 	Int. J. Mod. Phys. A 28, 1350012 (2013);
  M. Faizal and D. J. Smith, Phys. Rev. D 85, 105007 (2012). 
  \bibitem{mf1}M. Faizal,	Int. J. Mod. Phys. A 28,  1350012 (2013); Grav. Cosmol. 20,  
   132  (2014); P. Weinreb and M. Faizal, Phys. Lett. B 748,  102 (2015). 
   \bibitem{mos}P. Yu. Moshin and A. A. Reshetnyak, Phys. Part.   Nucl. Lett.  14(2), 411  (2017); 	Int. J. Mod. Phys. A 31 (2016) 1650111; 	Phys. Lett. B 739 (2014) 110;
   Nucl. Phys. B 888 (2014) 92.
 \bibitem{bodo}B. Geyer, P. M. Lavrov and P. Yu. Moshin, Phys. Lett. B 463 (1999) 188.
  \bibitem{pri}F. Ullah, P. A. Ganai, C. P. Haritha and B. Mawlong,  Int. J. Geom. Meth. Mod. Phys. 15 (2018)  1850127; P. A. Ganai, M. A. Mir, I. Rafiqi  and N. U. Islam, Mod. Phys. Lett. A 32 (2017)  1750214.
 \bibitem{pri1}P. A. Ganai, M. B. Shah, M. Syed and O. Ahmad, Int. J. Theor. Phys. 57 (2018)  1974; M. B. Shah and P. A. Ganai, Int. J. Geom. Meth. Mod. Phys. 15 (2018)   1850106; Int. J. Geom. Meth. Mod. Phys. 15 (2017)  1850009;  
  Commun. Theor. Phys. 69 (2018)   166; S. Masood, M. B. Shah and P. A. Ganai, Int. J. Geom. Meth. Mod. Phys. 15 (2017)  1850021.
 
 
 
 
  \bibitem{13} K. Yokoyama, Prog. Theor. Phys. 51, 1956 (1974).
   
 \bibitem{14}  K. Yokoyama, Prog. Theor. Phys. 51, 1956 (1974).
   
   \bibitem{15}  K. Yokoyama and R. Kubo, Prog. Theor. Phys. 52, 290 (1974).
   
   \bibitem{16} K. Izawa, Prog. Theor. Phys. 88, 759 (1992).
   
   
   \bibitem{17}    M. Koseki, M. Sato, and R. Endo, Prog. Theor. Phys. 90, 1111 (1993).
    
   \bibitem{18} R. Endo, Prog. Theor. Phys. 90, 1121 (1993).
    
   \bibitem{19} T. Saito, R. Endo, and H. Miura, Prog. Theor. Exp. Phys.  023B02 (2016).
    
   \bibitem{20}  K. Yokoyama, Prog. Theor. Phys. 59, 1699 (1978).
    
   \bibitem{21}   K. Yokoyama, M. Takeda, and M. Monda, Prog. Theor. Phys. 60, 927 (1978).
    
   \bibitem{22}  K. Yokoyama, Prog. Theor. Phys. 60, 1167 (1978).
    
   \bibitem{23} K. Yokoyama, Phys. Lett. B 79, 79 (1978).
   
   \bibitem{24} M. Faizal, Commun. Theor. Phys.57, 637 (2012).
    
   \bibitem{25}   M. Faizal, Mod. Phys. Lett. A 27, 1250147 (2012).
 
   \bibitem{26}  S. Upadhyay, Ann. Phys. 344, 290 (2014).
  
   \bibitem{27}  S. Upadhyay, Eur. Phys. J. C 74, 2737 (2014).
    
\bibitem{ao}M. Aochi, R. Endo   and H. Miura, Prog. Theor. Exp. Phys.  023B03 (2018).
\bibitem{faiz}M. Faizal, B. P. Mandal and S. Upadhyay,   Phys. Lett. B 721, 159 (2013).
 \bibitem{faiz1} M. Faizal, S. Upadhyay and B. P. Mandal, Phys.
Lett. B 738, 201 (2014);  	Int. J. Mod. Phys. A 30, 1550032 (2015).
\bibitem{faiz2}  M. Faizal, S. Upadhyay and B. P. Mandal,   Eur.
Phys. J. C 76,  189 (2016).
\bibitem{sudu}P. Yu. Moshin, S. Upadhyay and R. A. Castro, Braz. J. Phys. 47 (2017) 411.
\bibitem{pr}S. Upadhyay and P. A. Ganai, 	Prog. Theor. Exp. Phys. 2016, 063B04. 
\bibitem{suf} M. Faizal, S. Upadhyay  and B. P. Mandal,  Eur. Phys. J. C 76 (2016) 189.
\bibitem{suf1} S. Upadhyay  and B. P. Mandal, Mod. Phys. Lett. A25, 3347 (2010);
S. Upadhyay, S. K. Rai and B. P. Mandal, J. Math. Phys. 52, 022301 (2011).
 \bibitem{sudb} S. Upadhyay  and B. P. Mandal, Prog. Theor. Exp. Phys. 053B04 (2014). 
 \bibitem{sudu1} S. Upadhyay, Europhys. Lett. 105, 21001 (2014); 	Int. J. Theor. Phys. 55, 4005 (2016).
\bibitem{sud}S. Upadhyay, Eur. Phys. J. C  75, 593 (2015). 
 
\bibitem{Froissart} M. Froissart, Nuovo Cim. Suppl. {{14}, 197 (1959).}

\end{thebibliography}
\end{document}